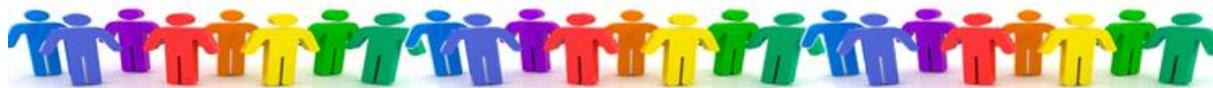

# When to Adjust Alpha During Multiple Testing: A Consideration of Disjunction, Conjunction, and Individual Testing

Mark Rubin
*The University of Newcastle, Australia*

## Abstract

Scientists often adjust their significance threshold (alpha level) during null hypothesis significance testing in order to take into account multiple testing and multiple comparisons. This alpha adjustment has become particularly relevant in the context of the replication crisis in science. The present article considers the conditions in which this alpha adjustment is appropriate and the conditions in which it is inappropriate. A distinction is drawn between three types of multiple testing: *disjunction testing*, *conjunction testing,* and *individual testing*. It is argued that alpha adjustment is only appropriate in the case of disjunction testing, in which *at least one* test result must be significant in order to reject the associated joint null hypothesis. Alpha adjustment is inappropriate in the case of conjunction testing, in which *all* relevant results must be significant in order to reject the joint null hypothesis. Alpha adjustment is also inappropriate in the case of individual testing, in which *each* individual result must be significant in order to reject each associated individual null hypothesis. The conditions under which each of these three types of multiple testing is warranted are examined. It is concluded that researchers should not automatically (mindlessly) assume that alpha adjustment is necessary during multiple testing. Illustrations are provided in relation to joint studywise hypotheses and joint multiway ANOVAwise hypotheses.

Correspondence concerning this article should be addressed to Mark Rubin at the School of Psychology, Behavioural Sciences Building, The University of Newcastle, Callaghan, NSW 2308, Australia. E-mail: Mark-Rubin@outlook.com  Web: http://bit.ly/rubinpsyc

The multiple testing of hypotheses occurs in the most areas of science. For example, it occurs in clinical science, where researchers investigate whether a treatment affects multiple disease symptoms, and it occurs in psychology, where researchers investigate whether multiple groups of people hold different attitudes to one another.

Multiple testing has been implicated in the replication crisis in science (e.g., Benjamin et al., 2018; Forstmeier et al., 2017; Goodman et al., 2016). In particular, it has been suggested that researchers who do not adequately correct their significance threshold, or *alpha level*, during multiple testing are at a greater risk of making Type I errors (incorrectly rejecting null hypotheses) and, consequently, publishing nonreplicable false positive results (Goodman et al., 2016).

Many books and articles explain *how* to adjust alpha levels during multiple testing (e.g., Bretz et al., 2011; Dmitrienko & D'Agostino, 2013; Dudoit & Van der Laan, 2008; Goeman & Solari, 2014; Hsu, 1996; Klockars, 2003; Pan, 2013; Shaffer, 1995; Streiner, 2015). However, far fewer articles consider *when* to adjust alpha levels during multiple testing (Proschan & Waclawiw, 2000). The most common view is that alpha adjustment is almost always required during multiple testing. For example, in their article on the control of false positives in neuroimaging, Bennett et al. (2009, p. 417) explained that "it is a statistical necessity that we must adapt our threshold criteria to the number of statistical tests completed on the same dataset." Similarly, in their tutorial on multiple testing in genomics, Goeman and Solari (2014) argued that "there can be no reason not to correct for multiple testing in a genomics experiment" (p. 24). Commentators who hold different views tend to be in complete opposition to any alpha adjustment. For example, O'Keefe (2003, p. 431) argued that "the practice of requiring or employing such adjustments should be abandoned," and Rothman (1990, p. 43) argued that "a policy of not making adjustments for multiple comparisons is preferable because it will lead to fewer errors of interpretation" (see also Hurlbert & Lombardi, 2012, p. 30; Mead, 1988; Perneger, 1998; Rothman et al., 2008; Sinclair et al., 2013; Stewart-Oaten, 1995; Wilson, 1962; for a brief review, see Hurlbert & Lombardi, 2012, pp. 30-31). Researchers cannot be blamed for being confused about alpha adjustment when they are confronted with these contradictory viewpoints.

Some articles have provided a more moderate and nuanced perspective in which an alpha adjustment is warranted in some cases of multiple testing but not in others (Armstrong, 2014; Bender & Lange, 2001; Greenland, 2020; Hewes, 2003; Matsunaga, 2007; Proschan & Waclawiw, 2000; Schulz & Grimes, 2005; Streiner, 2015; Tutzauer, 2003; Wason et al., 2014; Weber, 2007). Although numerous qualifying conditions have been proposed, a common criterion relates to the distinction between exploratory and confirmatory research. Some researchers believe that alpha adjustment is more appropriate when multiple testing occurs in exploratory research situations that involve unplanned analyses rather than in confirmatory research situations that include planned analyses (e.g., Armstrong, 2014; Cramer et al., 2016; Streiner, 2015; for a review, see Frane, 2015). However, other researchers hold the opposite view – that alpha adjustment is more appropriate in confirmatory situations than in exploratory situations (e.g., Bender & Lange, 2001; Schochet, 2009; Stacey et al., 2012; Tutzauer, 2003; Wason et al., 2014; for a discussion, see Parker & Weir, 2020, p. 3). Hence, the distinction between exploratory and confirmatory research does not seem to clarify when to adjust alpha.

In the present article, I consider an alternative approach to determining when to adjust alpha during multiple testing. Rather than being based on the type of research situation (exploratory vs. confirmatory), my approach is based on the type of multiple testing. Specifically, I consider three types of multiple testing – *disjunction testing, conjunction testing*, and *individual testing*. I argue that an alpha correction for multiple testing is only necessary in the case of disjunction testing and not in the cases of either conjunction or individual testing. I explain when it is appropriate to undertake each type of multiple testing and, consequently, when it is appropriate to adjust alpha. Based on this explanation, I argue that researchers should not automatically assume that alpha adjustment is necessary during multiple testing. I provide illustrations of the problems with this automatic (mindless) alpha adjustment assumption in relation to joint studywise hypotheses and joint multiway ANOVAwise hypotheses. I begin with an introduction to the issue of multiple testing.

## What is Multiple Testing?

To understand multiple testing, it is first necessary to understand the null hypothesis significance testing approach. This approach is based on *p* values. A *p* value is the probability of obtaining a test statistic value or a more extreme value in a sample assuming that (a) the sample was drawn from a null population, as described in the null hypothesis, and that (b) all statistical assumptions are valid. In order to decide whether a test result is "significant," researchers judge their observed *p* value against a threshold criterion value or alpha level.[1] If the *p* value for an observed test statistic is less than or equal to the alpha level, then researchers categorize the result as being "significant," and they decide to provisionally reject the null hypothesis that the sample was drawn from the null population. Otherwise, they categorise the observed result as "nonsignificant" and retain the null hypothesis.

In many fields, researchers set their alpha level at .05, meaning that they are willing to accept that random measurement error and random sampling error will cause them to incorrectly reject the null hypothesis in no more 5.00% of a long run of exact replications of their test. Hence, there is a 5.00% probability that researchers will make a Type I error in the long run by rejecting the null hypothesis when it is true.

It should be noted that null hypothesis significance testing is a hybrid of the Fisherian and Neyman-Pearson approaches (Dennis et al., 2019; Rubin, 2021). A key difference between these two approaches is that the Neyman-Pearson approach explicitly contrasts the null hypothesis with a formal alternative hypothesis, whereas the Fisherian approach does not. In addition, some neo-Fisherians do not use significance thresholds to make dichotomous "reject" vs. "fail to reject" decisions about the null hypothesis (Rubin, 2021, Footnote 4). However, many Fisherians, including Fisher himself, do use significance thresholds to make such decisions, and the issue of multiple testing is relevant to them (e.g., Fisher, 1971, pp. 205-207).

Imagine a case in which the same hypothesis is tested twice. For example, imagine that a group of researchers investigate the alternative hypothesis that eating jelly beans causes acne (Munroe, 2011). There are many different colours of jelly bean and so, to keep their study simple, the researchers randomly select two colours for testing: green and red. The researchers ask one group of participants to eat a bag of green jelly beans every day for a week and one group to eat a bag of red jelly beans every day for a week. The researchers also ask a control group of participants to eat a bag of sugar pills every day for a week. The researchers then count the number of spots on participants' faces.

In this jelly beans study, the researchers can make *multiple comparisons* in order to test the null hypothesis that the amount of acne among people who eat jelly beans is no greater than the amount of acne among people who eat sugar pills. In particular, the researchers can test for a significant increase in acne between (a) the green jelly beans group and the control (sugar pills) group and (b) the red jelly beans group and the control group. Hence, the researchers are conducting two tests of the same null hypothesis that eating jelly beans does not cause acne.[2]

Note that a hypothesis that undergoes multiple testing is called a *joint hypothesis.* Joint hypotheses comprise two or more *constituent hypotheses*. Hence, in the above example, the joint alternative hypothesis is that "eating jelly beans causes acne," and the constituent alternative hypotheses are that (a) "eating green jelly beans causes acne," and (b) "eating red jelly beans causes acne."

Imagine that the researchers in the jelly beans study use an alpha level of .05 as the significance threshold for their two one-sided tests. Further imagine that they find that the comparison between the green jelly beans group and the control group results in a significant *p* value of .030, but that the comparison between the red jelly beans group and the control group results in a nonsignificant *p* value of .070. What decision should the researchers make about the joint null hypothesis that eating jelly beans does not increase acne? There are three main approaches that they could take.

First, the researchers could require that *at least one* of the two tests returns a significant result before they reject the joint null hypothesis. This "at-least-one-test-significant" strategy (Dmitrienko & D'Agostino, 2013) represents a *disjunction testing* approach, because it operates on the basis of a logical disjunction decision rule (Weber, 2007).

Second, the researchers could require that *both* tests return a significant result before they reject the joint null hypothesis. This "all-tests-significant" strategy represents a *conjunction testing* approach, because it operates on the basis of a logical conjunction decision rule (e.g., Capizzi & Zhang, 1996; Dmitrienko & D'Agostino, 2013; Weber, 2007).

Finally, the researchers could abstain from making a decision about the joint null hypothesis and instead only make decisions about each of the two constituent null hypotheses. For example, this *individual testing* approach might allow the researchers to conclude that eating red jelly beans causes acne, but eating green jelly beans does not.

Below, I discuss each of these three types of multiple testing and their implications for adjustments to the alpha level. I illustrate my discussion with examples from psychology, clinical science, genomics, and neuroimaging in order to show how scientists might benefit from these different approaches to multiple testing.

## Disjunction, Conjunction, and Individual Types of Multiple Testing
### Disjunction Testing

Disjunction testing is also called *union-intersection testing* (Bretz et al., 2011, p. 20; Hochberg & Tamrane, 1987, p. 28; Kim et al., 2004; Parker & Weir, 2020, p. 2; Roy, 1953), because multiple constituent alternative hypotheses form a union (dotted area in Figure 1), and multiple constituent null hypotheses form an intersection (grey area in Figure 1).

Because the constituent null hypotheses form an intersection, it is only necessary to reject *one* of them in order to reject the corresponding joint intersection null hypothesis. For example, it is only necessary to reject the constituent null hypothesis that "green jelly beans do not cause acne" in order to reject the joint null hypothesis that "*neither* green jelly beans *nor* red jelly beans cause acne" and infer that "eating (either green or red) jelly beans causes acne."

Importantly, disjunction testing increases the probability of making a Type I error about the joint intersection null hypothesis, because it increases the number of opportunities that researchers have to incorrectly reject this hypothesis. In particular, if researchers undertake disjunction testing, then every test of a constituent hypothesis represents an opportunity to reject the joint null hypothesis. Consequently, when undertaking disjunction testing, it is important to know the probability of making *at least one* Type I error in the collection, or *family*, of constituent null hypotheses. This Type I error rate is called the *familywise error rate*.

*Figure 1.* Illustration of disjunction testing. Based on Kim et al. (2004, Figure 1[a]).

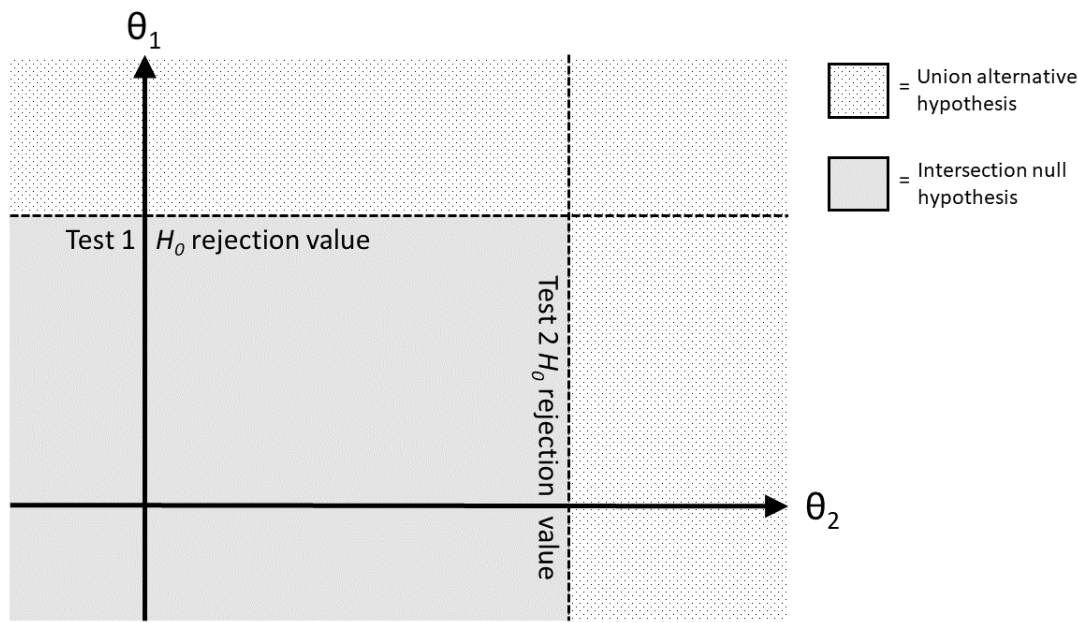

Assuming that test results are independent from one another, the familywise error rate is computed by determining the probability that at least one of the tests of the constituent null hypotheses in the family is significant when the joint null hypothesis is true. This probability is equal to 1.00 – (the probability that none of the tests are significant). Following the multiplicative probability rule for independent events, the probability that none of the tests are significant when the joint null hypothesis is true is equal to the product of the probabilities that each of them is nonsignificant (i.e., 1 - α). Hence, for *k* constituent null hypotheses that are each tested using an alpha level of α, the familywise error rate is equal to $1 – (1 - α)^k$. For example, the probability that at least one of two tests will result in a Type I error at the .05 alpha level is equal to $1.00 – (1 - .05)^2 = .098$.[3] Note that this Type I error rate is higher than the prespecified alpha level of .05. Consequently, if researchers use a disjunction testing approach, and they wish to maintain the probability of making a Type I error about the joint null hypothesis at the conventional alpha level (i.e., $α_{Joint} = .05$), then they need to decrease the alpha level for each constituent null hypothesis (i.e., $α_{Constituent} < α_{Joint}$).[4]

The amount by which $α_{Constituent}$ needs to be decreased can be determined using an alpha adjustment approach. There are many different alpha adjustment approaches (e.g., the Benjamini-Hochberg, Bonferroni, Dunn-Šidák, Holm, and Hochberg corrections; for a review, see Goeman & Solari, 2014). For example, the Dunn-Šidák correction uses the formula $1 – (1 - α)^{1/k}$ (Šidák, 1967). If this correction is used in the case of two constituent null hypotheses, then $α_{Constituent}$ should be reduced from .050 to .025 in order to maintain the Type I error rate for the joint null hypothesis at the prespecified $α_{Joint}$ of .050.

The familywise error rate can be contrasted with the false discovery rate, which is the expected proportion of incorrectly rejected null hypotheses (Benjamini & Hochberg, 1995, p. 290). If all of the null hypotheses are true, then the false discovery rate is equivalent to the familywise error rate. However, if some of the null hypotheses are false, then the false discovery rate will be less than the familywise error rate, because the false null hypotheses that are rejected do not count as erroneous rejections. Hence, unlike the familywise error rate, the false discovery rate is not conditioned on the joint null hypothesis being true, because it assumes that some of the associated constituent hypotheses may be false and, consequently, that the joint null hypotheses may be false.

**Conjunction Testing**

Disjunction testing represents an "at-least-one-test-significant" approach to joint null hypothesis testing. In contrast, conjunction testing represents an "all-tests-significant" approach. Berger (1982) proposed this approach as an *intersection-union test* (Berger, 1982; Berger & Hsu, 1996; Bretz et al., 2011, p. 22). The intersection-union test refers to a configuration of multiple constituent alternative hypotheses as an intersection (dotted area in Figure 2) and multiple constituent null hypotheses as a union (grey area in Figure 2).

*Figure 2.* Illustration of conjunction testing. Based on Kim et al. (2004, Figure 1[b]).

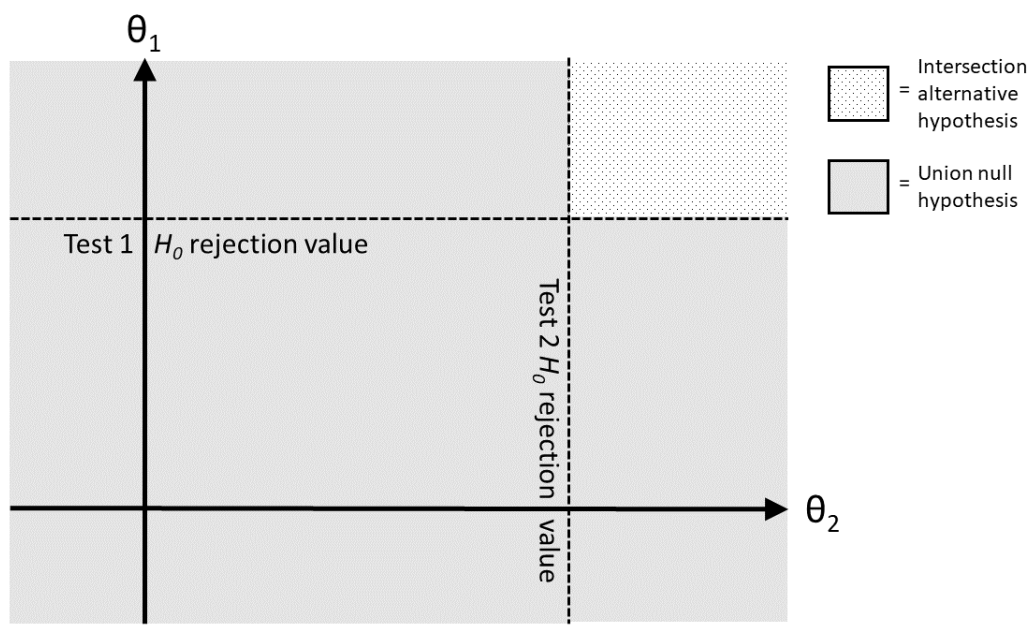

Because the constituent null hypotheses form a union, it is necessary to reject *all* of them in order to reject the corresponding joint union null hypothesis (grey area). For example, it is necessary to reject *both* the null hypothesis that "green jelly beans do not cause acne" *and* the null hypothesis that "red jelly beans do not cause acne" in order to reject the joint union null hypothesis that "*either* green jelly beans do not cause acne *or* red jelly beans do not cause acne" and infer that "all (tested) jelly beans cause acne."

A key aspect of conjunction testing is that it does not require an adjustment to the alpha level for tests of each constituent null hypothesis (i.e., $\alpha_{Constituent} = \alpha_{Joint}$; Berger, 1982; Dmitrienko et al., 2009; Dmitrienko & D'Agostino, 2013; Kim et al., 2004; Kordzakhia et al., 2010; Mascha & Turan, 2012; Massaro, 2009; Neuhäuser, 2006; Pan, 2013; Rubin, 2017b; Weber, 2007; Westfall et al., 2001; Winkler et al., 2016). This is because, although researchers use multiple tests to test the joint union null hypothesis, they may only reject this hypothesis if and only if *all* of their tests yield significant results. Hence, they only have a *single* opportunity to reject the joint null hypothesis at its prespecified alpha level of $\alpha_{Joint}$ (Mascha & Turan, 2012). Consequently, $\alpha_{Constituent}$ does not need to be reduced to compensate for multiple testing.[5]

One disadvantage of conjunction tests is that they lack statistical power, because they fail to reject the joint null hypothesis if one or more constituent tests yield a nonsignificant result (Francis & Thunell, 2021; Julious & McIntyre, 2012). For example, imagine that a researcher wants to undertake a conjunction test with a power level of .80 (i.e., $\beta_{Joint}$ = .20). If two constituent hypotheses are each tested with a power value of .80 (i.e., $\beta_{Constituent}$ = .20), then the familywise *Type II* error rate will be $1.00 - (1 - .20)^2 = .36$, which is higher than the $\beta_{Joint}$ of .20. This Type II error rate of .36 equates to a power level of .64, which is lower than the desired power of .80.

Conjunction testing is relatively common in clinical and translational science, in which treatments need to be shown to be effective on multiple aspects of a disease in order to be regarded as being successful (Dmitrienko et al., 2009; Dmitrienko & D'Agostino, 2013; Julious & McIntyre, 2012; Kordzakhia et al., 2010; Mascha & Turan, 2012; Massaro, 2009; Neuhäuser, 2006; Pan, 2013; Westfall et al., 2001). For example, researchers may test a new therapy for Alzheimer's disease by requiring it to be effective on both cognition and global clinical scores (Dmitrienko et al., 2009; Dmitrienko & D'Agostino, 2013). Similarly, clinical treatments for chronic obstructive pulmonary disease are usually required to demonstrate both (a) improved forced expiratory volume and (b) symptomatic benefits (Neuhäuser, 2006). Conjunction testing may also be used to test the effectiveness of combination therapies such as exercise and diet to control weight gain; antihistamine and decongestant to treat allergic rhinitis; and bronchodilators and inhaled corticosteroids to treat asthma (Westfall et al., 2001).

Conjunction testing has also been used in comparative genomics. Here, researchers are interested in identifying the same instances of gene expression in different species in order to draw conclusions about the generality of molecular or developmental mechanisms that underlie processes such as aging, energy metabolism, and diseases (Kim et al., 2004). For example, researchers may use conjunction testing to identify genes that are differentially expressed in the same way in response to caloric restriction in fruit flies, nematodes, *and* mice (for a worked example, see Kim et al., 2004).

Finally, conjunction testing has been used in neuroimaging (e.g., Nichols et al., 2005). Here, researchers might compare differences between several task groups and a control group in order to determine differences in the activation of thousands of voxels, each of which represent different parts of an image of the brain. Conjunction testing has been employed in order to confirm that certain brain regions are active under two or more different tasks (Nichols et al., 2005; Winkler et al., 2016).

**Individual Testing**

Disjunction and conjunction testing allow researchers to test a joint null hypothesis that comprises two or more constituent null hypotheses. In contrast, individual testing only allows researchers to test individual null hypotheses that do not comprise a joint null hypothesis. Hence, individual testing allows decisions about individual null hypotheses but not about joint null hypotheses. For example, in the jelly bean study, individual testing would allow the researchers to infer that eating green jelly beans causes acne, but it would not allow researchers to infer that eating jelly beans in general causes acne. Consequently, individual testing is most appropriate when researchers are not interested in testing joint null hypotheses.

Like conjunction testing, individual testing does not require an adjustment to the alpha level of each test ($\alpha_{Individual}$; Armstrong, 2014, p. 505; Cook & Farewell, 1996, pp. 96-97; Fisher,

1971, p. 206; Hewes, 2003, p. 450; Hurlbert & Lombardi, 2012, p. 30; Matsunaga, 2007. p. 255; Parker & Weir, 2020, p. 2; Rothman, 1990, p. 45; Rubin, 2017b, pp. 271-272; Rubin, 2020, p. 380; Savitz & Olshan, 1995, p. 906; Senn, 2007, p. 150; Sinclair et al., 2013, p. 19; Tukey, 1953, p. 82; Turkheimer et al., 2004, p. 727; Veazie, 2006, p. 809; Wilson, 1962, p. 299). This point is often misunderstood (e.g., O'Keefe, 2003) and so it is important to clarify it. If multiple test results are used to make a decision about a single joint null hypothesis, and disjunction testing is used, then each test represents an independent opportunity to reject the joint null hypothesis, and the alpha level of each test ($\alpha_{Constituent}$) needs to be lowered in order to compensate for the increased number of opportunities to make a Type I error about the joint null hypothesis. In contrast, if a single test result is used to make a decision about a single null hypothesis, then that test result provides only one opportunity to make a Type I error about that null hypothesis. Consequently, the alpha level of the test ($\alpha_{Individual}$) does not need to be lowered.

Importantly, the logic of individual testing applies even when multiple instances of individual testing take place side-by-side within the same study (see also Cook & Farewell, 1996; Fisher, 1971, p. 206; Greenland, 2020, p. 5; Hurlbert & Lombardi, 2012, p. 30; Kotzen, 2013; Parker & Weir, 2020, p. 2; Rubin, 2017b, pp. 271-272; Savitz & Olshan, 1995, p. 906; Senn, 2007, p. 150; Tukey, 1953, pp. 82-83; Wilson, 1962). If each decision to reject each individual null hypothesis depends on no more than one significance test, then none of the individual tests constitute a "family" with respect to any single hypothesis. Consequently, it is not necessary to adjust alpha levels on the basis of any family-based error rate (e.g., familywise error rate, per family error rate, etc.; Hurlbert & Lombardi, 2012, p. 30). A family-based alpha adjustment is only necessary when researchers undertake disjunction testing of a joint intersection null hypothesis.

Of course, a researcher who conducts a greater number of individual tests will have a greater opportunity to obtain more significant results and, consequently, a greater opportunity to make more Type I errors (e.g., Drachman, 2012; Goeman & Solari, 2014). For example, imagine that a researcher tests 100 *true* null hypotheses using 100 individual tests that each have an $\alpha_{Individual}$ of .05. In this case, the researcher has a greater opportunity to obtain more significant results and make more Type I errors than if they had only tested one true null hypothesis. Indeed, given that all 100 null hypotheses are true, the researcher should expect to obtain five significant results and, consequently, make five Type I errors. However, it is important not to confuse this expected outcome for the collection of individual tests (the *per family error rate*) with the probability of making a Type I error in relation to *each* individual test (the *individual*, *marginal*, or *per determination* error rate; Cook & Farewell, 1996, pp. 96-97; Tukey, 1953, p. 82). As the size of the family of tests increases, the individual error rate remains constant (i.e., $\alpha_{Individual} = .05$; Senn, 2007, pp. 150-151). It is only the per family error rate that increases (i.e., $\alpha \times k_{family}$).

To illustrate, if a person rolls a 20-sided dice 20 times instead of once, then they will increase the familywise probability that they will roll a "3" in at least one of their rolls from .05 to .64. However, they will not increase the individual probability that each roll will result in a "3." This individual probability will always remain at .05, regardless of the number of rolls of the dice (for a similar example, see Kotzen, 2013). Hence, it is perfectly true that "the more tests that are run, the greater the likelihood that at least 1 will be significant by chance" (Streiner, 2015, p. 722). However, if researchers undertake individual testing using an $\alpha_{Individual}$ of .05, then it is also true that the probability that they will make a Type I error *in the case of each specific individual hypothesis test* is no more than 5.00%. It is a form of gambler's fallacy to believe that each successive individual test in a series of individual tests has a greater than 5.00% chance of yielding a Type I error, even after the millionth test.

The uncomfortable feeling that some researchers might feel about conducting multiple individual tests may be attributed to a confusion between the alpha levels that are associated with individual testing ($\alpha_{Individual}$) and the alpha levels that are associated with disjunction testing ($\alpha_{Constituent}$). To illustrate, consider the jelly beans study again, as originally conceived by Munroe (2011) in Figure 3.

Munroe's (2011) jelly bean study is supposed to highlight the inappropriateness of not adjusting the alpha level during multiple testing. However, it actually illustrates the confusion between $\alpha_{Individual}$ and $\alpha_{Constituent}$ in a case of individual testing. In the study, the scientists conducted individual tests of 20 different hypotheses (i.e., one test per hypothesis), and they obtained a single significant result using an alpha level of $\alpha_{Individual} = .05$. Based on the results of these individual tests, they inferred that there is "a link between green jelly beans and acne." Contrary to Munroe's intimation, this inference is entirely appropriate given its level of specificity – it refers to *green* jelly beans only and not to jelly beans of one or more unspecified colours – and the fact that it is based on a single significance test that used a conventional alpha level of .05 (for the same conclusion, see Lew, 2019, pp. 21-22). Hence, in this case, there is no more than a 5.00% probability that the scientists' decision to reject the associated null hypothesis (i.e., "green jelly beans do not cause acne") represents a Type I error.[6]

The confusion in the jelly bean study relates to the fact that the scientists also have the *potential* to subsume their 20 hypotheses under a joint union alternative hypothesis that "either green, purple, brown, pink, blue, teal, salmon, red, turquoise, magenta, yellow, grey, tan, cyan, mauve, beige, lilac, black, peach, or orange jelly beans cause acne." As shown in Table 1, *if* they undertook disjunction testing of the corresponding joint intersection null hypothesis using an $\alpha_{Constituent}$ of .05 for each of the 20 constituent hypotheses, then the single significant result that they obtained would be likely to represent a Type I error in relation to the joint null hypothesis, because 1 out of every 20 significant results is expected to represent a Type I error when using an $\alpha_{Constituent}$ of .05 (i.e., .05 x 20 = 1; the *per family error rate*).

Importantly, if the scientists subsumed their 20 hypotheses under the joint union alternative hypothesis that "jelly beans (of one or more colours) cause acne," then their inference should be that "jelly beans (of one or more colours) cause acne." This inference would be inappropriate, because the scientists have a 64.15% probability of incorrectly rejecting the associated joint intersection null hypothesis when investigating 20 different colour of jelly bean. However, the scientists did not make this broader inference. Instead, they made the more specific inference that "*green* jelly beans cause acne." This more specific inference is appropriate given that the scientists only have a 5.00% probability of incorrectly rejecting the associated null hypothesis (Lew, 2019, pp. 21-22).

The problem in jelly bean study and more generally is that it is easy to confuse the alpha level for each hypothesis test in the individual testing situation (i.e., $\alpha_{Individual}$) with the alpha level for each hypothesis test in the family testing situation (i.e., $\alpha_{Constituent}$) and to conclude that $\alpha_{Individual}$ needs to be adjusted because, *if the 20 tests formed a family*, then $\alpha_{Constituent}$ would need to be adjusted (see also Greenland, 2020, p. 5). This *alpha confusion* leads to the erroneous conclusion that a single significant result that is obtained following 20 individual tests that each use an $\alpha_{Individual}$ of .05 is more likely to be a Type I error than a single significant result that is obtained using a single individual test that uses an $\alpha_{Individual}$ of .05 (e.g., Feise, 2002; Sainani, 2009).

*Figure 3.* Illustration of multiple individual testing. Retrieved from https://xkcd.com/882/

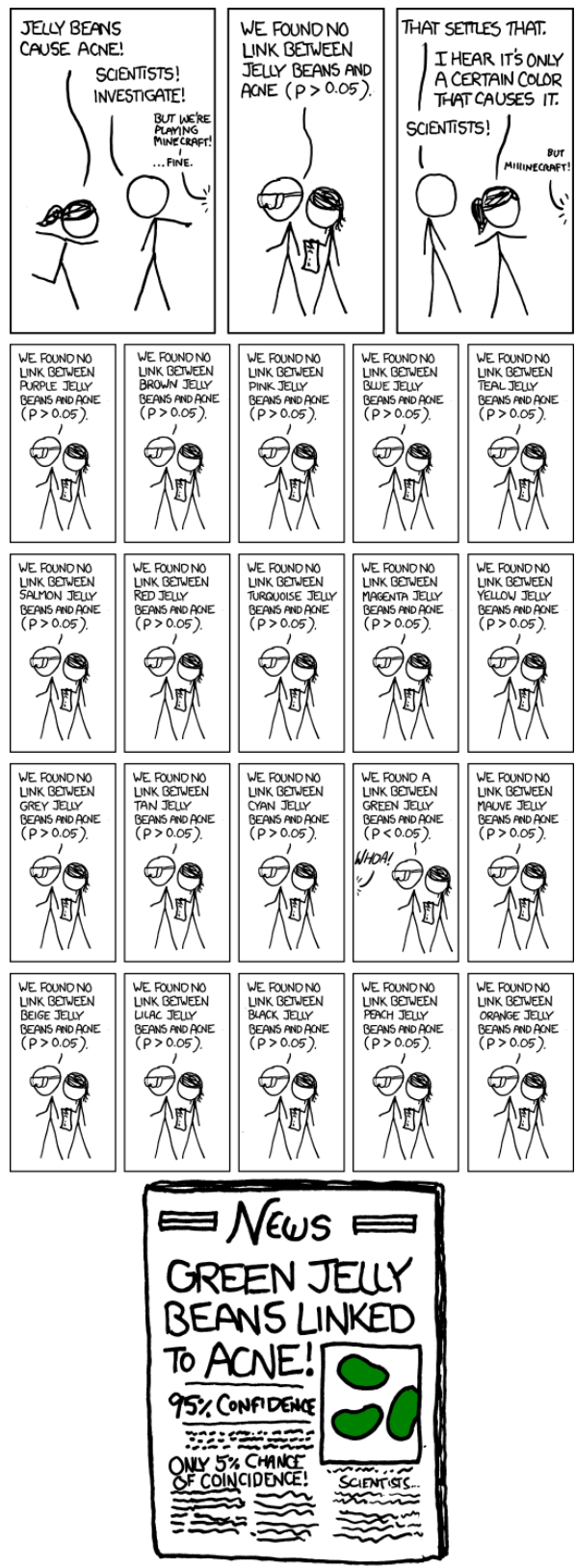

Table 1

*Distinguishing Type I Error Rates for Decisions About Joint and Individual Null Hypotheses*

| | A Single Decision About a Single Joint Intersection Null Hypothesis | Single Decisions About Multiple Individual Null Hypotheses |
|---|---|---|
| Primary null hypothesis or hypotheses | *Neither* green, purple, brown, pink, blue, teal, salmon, red, turquoise, magenta, yellow, grey, tan, cyan, mauve, beige, lilac, black, peach, *nor* orange jelly beans cause acne. | 1. Green jelly beans do not cause acne.<br>2. Purple jelly beans do not cause acne.<br>3. Brown jelly beans do not cause acne.<br>…<br>20. Orange jelly beans do not cause acne. |
| Significance tests used to test the primary null hypothesis or hypotheses | 1. Green jelly beans vs. control.<br>2. Purple jelly beans vs. control.<br>3. Brown jelly beans vs. control.<br>…<br>20. Orange jellybeans vs. control. | 1. Green jelly beans vs. control.<br>2. Purple jelly beans vs. control.<br>3. Brown jelly beans vs. control.<br>…<br>20. Orange jellybeans vs. control. |
| Number of significance tests ($t$) | 20 | 20 |
| Number of primary null hypotheses ($h$) | 1 | 20 |
| Number of significance tests per primary hypothesis ($k = t/h$) | 20 | 1 |
| Alpha level used for each significance test ($\alpha$) | .050 ($\alpha_{Constituent}$) | .050 ($\alpha_{Individual}$) |
| Number of false positives expected among the significance tests: The per family error rate ($k\alpha$) | 1.00 | .050 |
| Type I error rate for each primary hypothesis: The familywise error rate: $(1 - [1 - \alpha]^k)$ | .64 | .050 |

*Note.* The computation of the familywise error rate assumes that the 20 tests are independent. In reality, this may not be the case. In the case of positive dependence, the familywise error rate will be lower than .64.

To clarify, in the jelly bean study, $\alpha_{Individual}$ is the value for 20 independent alphas (each set at .05) that are associated with 20 individual hypotheses that are each tested only once (e.g., "green jelly beans cause acne," "red jelly beans cause acne," "purple jelly beans cause acne," etc.). Consequently, none of these 20 alphas need to be adjusted, because none of them are associated with disjunction testing. In contrast, $\alpha_{Constituent}$ is the value for 20 disjunction tests of the same joint hypothesis (e.g., "jelly beans cause acne") that is tested using an $\alpha_{Joint}$ of .05. Consequently, $\alpha_{Constituent}$ needs to be reduced (e.g., to .0025) in order to maintain $\alpha_{Joint}$ at .05.

Previous commentators have also attempted to clarify this alpha confusion. In particular, Matsunaga (2007, p. 255) explained that,

> if multiple $H_0$s are tested, inflation is of no concern because Type I errors are partitioned per $H_0$, each of which entails distinct alphas. If multiple tests are carried out within one $H_0$, however, overall Type I error rate for that $H_0$ becomes inflated and adjustment needs to be made (see also Rubin, 2017b, p. 272).

In summary, if researchers perform 20 tests and obtain only one significant result using an alpha of .05, then they will have a 64.15% chance of making a Type I error with respect to a joint null hypothesis that is disjunction tested but only a 5.00% chance of making a Type I error with respect to an individual null hypothesis that is individually tested. It is for this reason that there is no contradiction in the two claims made by the scientists in Munroe's (2011) cartoon (Lew, 2019, p. 21). The scientists may have sufficient evidence to make the specific claim that "we found a link between green jelly beans and acne ($p < 0.05$)" (Panel 17 in Figure 3) while lacking sufficient evidence to make the broader claim that this link extends to jelly beans of some unspecified colour or colours, and so they would need to concede that "we found no link between jelly beans and acne $p > 0.05$" (Panel 2 in Figure 3).

## Multiple Testing and Selection Bias

In their discussion of multiple testing in genomics, Goeman and Solari (2014) proposed that the individual testing of multiple individual hypotheses does not necessitate a multiple testing correction, because "without multiple testing correction the probability of a type I error in each individual hypothesis remains equal to α regardless of the number of hypotheses that have been tested" (p. 2). Hence, Goeman and Solari hold a similar view to the one discussed above. However, they also proposed that the individual testing of multiple individual hypotheses can lead to a *selection bias* on the part of researchers (e.g., Benjamini & Bogomolov, 2011; Cox, 1965). Specifically, researchers may select and report significant results and fail to report nonsignificant results. According to Goeman and Solari, "multiple testing methods aim to correct for this selection process" (p. 2). In contrast, I argue that a selection bias only necessitates an alpha adjustment when testing joint null hypotheses and not when testing individual null hypotheses.

To illustrate, consider the jelly bean study once again. If the researchers undertook individual testing, and they reported the significant result for green jelly beans without reporting the nonsignificant results for the other 19 colours of jelly bean, then the $\alpha_{Individual}$ level for the individual hypothesis that "green jelly beans cause acne" would remain valid, because a single test has been used to make a decision about a single individual hypothesis. Hence, the selection bias does not inflate the alpha level of individual tests during individual testing (for related discussions, see Kotzen, 2013, p. 167; Rubin, 2017a, p. 325; Rubin, 2017c; Rubin, 2017d, pp. 316-317; Rubin, 2020).[7]

In contrast, a selection bias *can* lead to alpha inflation when testing joint hypotheses. Imagine that the scientists undertook 20 disjunction tests of the joint null hypothesis that "jelly beans do not cause acne," but they retained their $\alpha_{Constituent}$ level at the conventional level of .05 instead of lowering it to compensate for their disjunction testing. Further imagine that the researchers found a single significant effect for green jelly beans using this unadjusted $\alpha_{Constituent}$ level. In this case, failing to report the results of the other 19 tests misrepresents the situation as one of individual testing rather than disjunction testing, and the researchers may incorrectly infer that "jelly beans cause acne" (i.e., a joint hypothesis) on the basis of an $\alpha_{Individual}$ of 5.00% when the actual Type I error rate for this inference is 64.15% (i.e., $\alpha_{Joint}$). Hence, the selection bias inflates the relevant alpha level when testing joint null hypotheses but not when testing individual null hypotheses.

More generally, selecting an effect from among a variety of other unrelated effects because it is larger than the other effects does not necessarily mean that the selected effect will be "biased." A bias will only occur when the selection occurs among different instances of the *same* effect, not when it occurs between qualitatively different effects. By analogy, picking the largest cherry from a bowl of cherries is likely to result in an unusually large cherry (i.e., a biased cherry). In contrast, selecting the largest fruit from a barrel that contains a variety of average-sized fruits is likely to yield an average-sized watermelon.

## Distinguishing Between Alpha Specification and Alpha Adjustment

None of the above points should be interpreted as suggesting that the alpha level during the individual testing of multiple hypotheses should always be set at the conventional .05 level. In every significance testing situation, researchers need to specify their alpha level in the context of a range of external background factors, including the plausibility of the hypothesis, the plausibility of potential alternative explanations, the theoretical and/or practical costs of Type I and Type II errors, the smallest effect size of interest, the sample size, and the variability in the data (Mudge et al., 2012). Hence, even in the individual testing situation, there may be grounds for lowering the alpha level below the conventional .05 threshold (e.g., Rothman et al., 2008, pp. 234-235). For example, a much lower alpha level would be appropriate when testing the implausible hypothesis that dead Atlantic salmon will exhibit brain activity in a specific brain region (Bennett et al., 2010), because extraordinary claims require extraordinary evidence. Importantly, this process of *alpha specification* is quite different from the previously discussed *alpha adjustment* during disjunction testing (for a similar view, see Parker & Weir, 2020, p. 564; Ryan, 1962, p. 305). In the former case, researchers specify an alpha level for their individual or joint hypothesis based on external factors. In the latter case, researchers adjust that prespecified alpha level in order to make it applicable to disjunction tests of constituent hypotheses.

In some cases, prudent alpha specification may be more appropriate than alpha adjustment. For example, in the field of genomics, researchers are interested in screening associations between hundreds of thousands of single nucleotide polymorphisms (SNPs) and diseases or other phenotypic traits in order to identify the largest and most reliable associations. Hence, they might attempt to identify the top 20 SNP associations among hundreds of thousands of tests (Goeman & Solari, 2014; Pan, 2013). In this case, there is a need to reduce the alpha level for each test not because researchers want to undertake disjunction testing of a genome-wide joint null hypothesis but because they want to achieve a more stringent screening approach in order to identify the largest effect sizes, which they presume are more likely to be clinically and biologically meaningful (Otani et al., 2018).

To illustrate, consider Wu et al.'s (2018) tests of associations between 167,355 SNPs from 532 pigs and phenotypic traits from the pigs' litters (e.g., number born). In order to maintain the genome-wide significance level at 5.00%, they used a Bonferroni correction (i.e., .05/167,355). On finding at least one single significant association, this Bonferroni correction would allow the researchers to reject the joint null hypothesis that the genome is not associated with the phenotype expression. However, the researchers did not make this genome-wide inference. Instead, they made SNP-specific inferences about "the top significant SNPs" (p. 173) and their associated chromosomes. For example, they noted that eight SNPs were significantly associated with the number of pigs born in a litter, that seven of these were located on the same chromosome, and that one had a novel location. Hence, the researchers adjusted their $\alpha_{Constituent}$ level, which enables a statistical inference about a genome-wide joint hypothesis, but they then ignored this joint hypothesis and instead made statistical inferences about individual hypotheses (i.e., which specific SNPs were associated with number of pigs born in a litter). Turkheimer et al.'s (2004) advice for functional brain imaging researchers is relevant here: "If before or after testing one wishes to consider the individual result on its own individual merit, then the multiple comparison correction becomes not only incorrect but also meaningless" (p. 727; see also Cook & Farewell, 1996; Cox, 1965). Genetics researchers have also commented on this inferential mismatch. For example, Otani et al. (2018) recently noted that "the FWER [familywise error rate] criterion strictly controls the probability of having at least one false positive in millions of tests, and geneticists should generally recognize its inappropriateness regarding the primary purposes of GWAS [genome-wide association studies]" (p. 1). According to Otani et al. (2018), the primary purpose of GWAS research is to identify SNPs that have comparably large effects, because these are most likely to be clinically and biologically meaningful. Given this purpose, it is more appropriate to use an individual testing approach in which the alpha level for each test has been specified at a more stringent level in order to screen out the smaller, less biologically important effects.

How are researchers supposed to determine a suitably stringent alpha level when they undertake multiple individual tests? As with single individual testing, a mix of community standards and cost analysis is required. In terms of community standards, conventional alpha levels can vary from field to field. For example, in a survey of 172 genome-wide association studies, Jannot et al. (2015) found that a consensus had emerged that an alpha level of .00000005 (i.e., $5.0 \times 10^{-8}$) is appropriate. In theory, this alpha level is based on a Bonferroni adjustment to the conventional 5.0% alpha level that assumes a million tests. However, in practice, it has been validated by considering the actual replicability of specific SNP-trait associations (Panagiotou et al., 2011). Hence, again, genomic researchers are more concerned about identifying specific SNP associations that are relatively large and replicable than they are about incorrectly rejecting the joint genome-wide null hypothesis. In terms of cost analysis, Type I errors need to be judged in relation to real world consequences and Type II errors. For example, Mudge et al. (2012) have proposed an optimal alpha approach that balances the costs of Type I and Type II errors in the context of a specified critical effect size (i.e., a smallest effect size of interest). In a meta-analysis of 242 microarray gene expression studies, Mudge et al. (2017) found that this optimal alpha approach resulted in Type I and II "error rates as low or lower than error rates obtained when using (i) no post-hoc adjustment, (ii) a Bonferroni adjustment and (iii) a false discovery rate (FDR) adjustment" (p. 1).

In summary, there is an important difference between using a million tests to identify the top 20 largest individual associations and using a million tests to disjunction test a joint intersection null hypothesis. A lower alpha level may be warranted in both cases. However, it is more appropriate to achieve this lower alpha through alpha specification in the former case (i.e., lower $α_{Individual}$ to screen out nonsignificant associations that are most likely below the smallest effect size of interest) and alpha adjustment in the latter case (i.e., $α_{Constituent} < α_{Joint}$ to maintain $α_{Joint}$ at .050).

### When Should Researchers Use Individual, Disjunction, and Conjunction Testing?

To recap, there are three approaches to multiple testing: disjunction testing, conjunction testing, and individual testing. Disjunction and conjunction testing allow researchers to test joint null hypotheses, but individual testing does not. Furthermore, disjunction testing requires an alpha adjustment, but conjunction and individual testing do not. Hence, in order to know when to adjust alpha, researchers need to know when to use each of these three types of multiple testing, and it is to this issue that I now turn.

The first point that researchers should consider is whether they are making a *statistical claim* that is warranted by a specific *p* value and alpha level. For example, based on the result of a *t* test and a conventional alpha level, a statistical claim might be: "Male participants had significantly higher self-esteem than female participants, $t(479) = 2.11$, $p = 0.018$." In contrast, more substantive non-statistical claims may summarise the results of significance tests without themselves being warranted by a specific *p* value (Meehl, 1978, p. 824). For example, a non-statistical claim might be: "Based on the results of Studies 1, 2, and 3, it was concluded that men have higher self-esteem than women." Note that this claim is not explicitly tied to a specific *p* value and alpha level. Importantly, the question of whether to adjust an alpha level only applies to statistical claims. This question does not apply to claims that are not tied to a specific *p* value, because such claims are not associated with a specific alpha level, and they may be in error due to not only random sampling and measurement error but also theoretical errors, model misspecification, systematic measurement error, and so on (Rubin, 2017b, p. 272).

If researchers *are* making a claim about statistical significance, then they need to consider whether their claim derives from the test of an individual null hypothesis or a joint null hypothesis. If they are testing an individual hypothesis, then they should use individual testing and an unadjusted alpha level (Cook & Farewell, 1996; Rothman et al., 2008, pp. 236-237; Wilson, 1962). If they are testing a joint hypothesis, then the decision about adjusting alpha depends on whether they are using disjunction testing or conjunction testing.

### Individual Hypotheses

Individual hypotheses are hypotheses than can be tested using a single significance test. In some cases, researchers' methods and designs constrain them into testing individual hypotheses. For example, researchers might have only one relevant predictor or comparison that relates to only one relevant outcome variable. Consequently, they have only one test that is relevant to their individual hypothesis. In this case, they are only able to conduct an individual test.

In other cases, researchers may have several predictor variables, comparison groups, and/or outcome variables. As discussed above, in these cases, researchers may undertake individual testing using an unadjusted alpha level in order to make separate decisions about each individual null hypothesis.

Researchers may also find that there are theoretical, practical, and/or empirical reasons (e.g., factor analyses) for aggregating across some of their constituent groups or variables in order to create composite groups or variables. They may then subject these composite groups or variables to individual testing at an unadjusted alpha level (Feise, 2002; Goeman & Solari, 2014; Hung &

Wang, 2010; Luck & Gaspelin, 2017; Matsunaga, 2007; Schulz & Grimes, 2005; Senn, 2007, p. 151).

A statistical aggregation approach may also be used to operationalize an individual test across groups or variables (Senn, 2007, p. 153). For example, a researcher might use a one-way ANOVA with simple contrasts that compare two experimental conditions to a control condition. Alternatively, a researcher might use a MANOVA to test the effects of a treatment on two or more outcome variables. To illustrate, consider the case of a clinical study that aimed to investigate the ability of a treatment to prevent premature infants from developing respiratory distress syndrome (RDS; Wang et al., 2015). There were three outcome variables: incidence of RDS at 24 hours, RDS-mortality through 14 days of age, and air leak through 7 days of age. As Dmitrienko (in Wang et al., 2015) explained, if it is necessary to demonstrate an effect of the treatment on a specific outcome (e.g., RDS-mortality) in order to mount the case for regulatory change, then disjunction testing would be inappropriate, because it would reject the joint null hypothesis on the basis of a significant result in relation to *any* of the three outcomes. Conjunction testing would be more appropriate in this case. However, it would lack power, which may be problematic in this particular scenario. Hence, Dmitrienko recommended using a single statistical test that provide a simultaneous assessment of the treatment effect across all three outcome variables and yields a single test statistic (e.g., a MANOVA).

**Joint Hypotheses**

The first requirement for testing a joint hypothesis is that the hypothesis should allow a statistical inference that has relevant and meaningful theoretical and/or practical implications (Cook & Farewell, 1996, p. 107; Cox, 1965, p. 223; Hochberg & Tamrane, 1987, p. 5; Parker & Weir, 2020, p. 2). To meet this requirement, researchers should ensure that the family of constituent hypotheses that comprise the joint hypothesis are theoretically consistent with their intended inference (see also Hung & Wang, 2010). In particular, the family must contain all relevant constituent hypotheses and no irrelevant constituent hypotheses (Cox, 1965; Hochberg & Tamrane, 1987, p. 6; Huberty & Morris, 1988, p. 572). It is helpful for researchers to make their research materials and data set publicly available online in order to allow others to verify the correct specification of their joint hypotheses and to check for any potential selection bias (Cox, 1965; Goeman & Solari, 2014; Rubin, 2017b, p. 273; Rubin, 2020).

The second requirement for testing a joint hypothesis is that researchers use an appropriate form of testing. Researchers should use disjunction testing when the rejection of *any* of the constituent hypotheses is sufficient to reject the joint hypothesis as a whole and the extent of generalisation across constituent hypotheses is unimportant. In contrast, researchers should undertake conjunction testing when it is important to demonstrate the confirmation of *all* constituent hypotheses within a joint hypothesis.

In the case of disjunction testing, researchers also need to assume that the constituent hypotheses are *theoretically exchangeable* with regards to inferences about the joint hypothesis under investigation (e.g., Rosset et al., 2018). That is to say, a significant result in relation to *any* of the constituent hypotheses must provide the same logical basis for rejecting the joint null hypothesis. For example, the hypothesis that "red M&Ms cause acne" is not theoretically exchangeable with the hypotheses that "green jelly beans cause acne" and "red jelly beans cause acne" when testing the joint hypothesis that "jelly beans cause acne," because M&Ms are not a type of jelly bean. Consequently, the red M&Ms hypothesis should not be included as a constituent hypothesis in the joint "jelly beans cause acne" hypothesis. Importantly, the exchangeability assumption is violated if researchers have an a priori theoretical expectation that one or more of their constituent hypotheses will yield a different result to the others. For example, if it is expected, a priori, that green jelly beans cause acne but that red jelly beans do not, then it would be inappropriate to include these two hypotheses as constituent hypotheses in the joint hypothesis that "jelly beans cause acne."

Conjunction testing may be more appropriate than disjunction testing when researchers undertake theory testing. Theories usually predict that *all* of their constituent hypotheses are true. They do not usually predict that *at least one* of their constituent hypotheses is true. Consequently, it is more logical for researchers to use conjunction testing rather than disjunction testing when they want to make a statistical inference about a joint hypothesis that comprises a family of hypotheses that belong to the same theory. Again, however, conjunction testing may suffer from lower power. In addition, theory evaluation may be better conceived as a "qualitative exercise," because it is influenced by non-statistical considerations (Haig, 2009, p. 220).

**Against an Automatic Alpha Adjustment Assumption**

To summarize, researchers only need to adjust their alpha level when they undertake disjunction testing of a joint null hypothesis. Furthermore, researchers should only undertake the disjunction testing of a joint null hypothesis when that hypothesis (a) enables a relevant theoretical and/or practical inference and (b) is better suited to disjunction testing rather than conjunction testing. This limited and qualified approach to alpha adjustment stands in contrast to the more common unqualified view that alpha adjustment is almost always necessary during multiple testing (e.g., Bennett et al., 2009; de Groot, 2014; Glickman et al., 2014). For example, in the introduction to their article on the false discovery rate, Glickman et al. provided the following explanation for alpha adjustment:

> The usual argument to convince researchers that adjustments are necessary when multiple tests are performed is to point out that, without adjustments, the probability of at least one null hypothesis being rejected is larger than acceptable levels. Suppose, for example, that a researcher performs 100 tests at the α = 0.05 significance level in which the null hypothesis is true in every case. If all the tests are independent, then the probability that at least one test would be incorrectly rejected is $1 - (1 - 0.05)^{100} = 0.9941$, or 99.41% (p. 851).

Similarly, in their article on multiple testing, Sainani (2009) provided the following explanation:

> Mathematically, the problem of multiple testing can be explained as follows: every statistical test comes with an inherent false positive, or type I error, rate—which is equal to the threshold set for statistical significance, generally .05. However, this is just the error rate for one test; when more than one test is run, the overall type I error rate is much greater than 5%. For example, if one runs 100 independent statistical tests where it is known no effects exist, the chance of getting at least one false positive (ie, at least one P value less than .05) is 99.4%...and 5 false positives are expected (because approximately 1 in 20 tests will yield a false positive) (p. 1089).

At this stage, the missing qualifications to these explanations should be apparent: (a) They assume that none of the 100 tests represent individual tests of individual hypotheses. (b) They assume that the 100 tests form a coherent family of tests in relation to a theoretically- and/or practically-relevant joint hypothesis. (c) They assume that researchers are undertaking a disjunction test of this joint hypothesis, rather than a conjunction test. To be clear, I am not suggesting that these three qualifications are never met. I am only suggesting that they are often ignored, as in the above examples, and that this omission leads to an inaccurate view that, when undertaking multiple testing, it is always necessary to compute family-based error rates and adjust alpha levels on the basis of these error rates.

It is important to note that automatic (mindless) alpha adjustment is not advocated by some of the experts in the field of multiple testing (Tukey, 1953, p. 82-83; see also Mead, 1988, pp. 310-314; Parker & Weir, 2020, p. 4). Instead, they argue that the choice between individual and disjunction testing should depend on the number and type of inferences that are to be made. If multiple testing is used to make multiple independent statistical inferences, then no alpha adjustment is warranted. Below, I illustrate the problems with automatic alpha adjustment in relation to *studywise error rates* and *multiway ANOVAwise error rates*.[8]

**Studywise Error Rates**

I use the term *studywise error rates* (sometimes called *experimentwise, global, or universal error rates*) to refer to family-based error rates (e.g., familywise error rates, per family error rates, false discovery rates, etc.) that are associated with all of the hypotheses that are tested in a study, experiment, or sample or, in the case of exploratory analyses, all of the hypotheses that could have been tested (e.g., An et al., 2013, pp. 6-7; Cohen, 1990, p. 1304; Drachman, 2012, p. 2, p. 2; Klockars, 2003, p. 614; Luck & Gaspelin, 2017, p. 151; Maxwell & Delaney, 2004, p. 291; Miller, 1981, p. 34; Parker & Weir, 2020, p. 3; Rubin, 2022; Ryan, 1962; Shaffer, 2006; Stacey et al., 2012, p. 1830). Consistent with the above points, researchers only need to consider the studywise error rate if they undertake disjunction testing of the joint studywise null hypothesis that the study produces a null effect. Furthermore, researchers should only be expected to test this joint studywise hypothesis if there are theoretical and/or practical reasons for doing so. However, often these reasons are lacking. As Cook and Farewell (1996, p. 106) explained with reference to clinical trials, "a concern is that testing strategies are frequently adopted with the aim of controlling the experimental type I error rate without considering how this relates to the questions of main interest." More recently, Parker and Weir (2020, p. 2) echoed this concern with respect to multi-arm clinical trials: "If treatments are distinct and we are interested in individual treatment versus control comparisons,…then it is difficult to see how the concept of formulating a global intersection null hypothesis could be relevant." If it is not useful to test the joint studywise hypothesis, then researchers should consider lower-order families of hypotheses and/or individual hypotheses for testing (Benjamini & Bogomolov, 2011; Efron, 2008; Fisher, 1971, p. 206; Hochberg & Tamrane, 1987, pp. 6-7; Hung & Wang, 2010; Mei et al., 2017; Rubin, 2017b). For example, in their discussion of multiple testing in microarray gene expression analysis, Yekutieli et al. (2006) explained that "the set of hypotheses that is of interest to the researcher in a single study does not necessarily form a single family of hypotheses" (p. 416). Instead, they suggested that families can be specified at the level of genes. Similarly, in discussing functional neuroimaging research, Benjamini and Bogomolov (2011) explained that hypotheses that refer to the same brain region should be regarded as belonging to the same family. Hence, joint studywise null hypotheses are often theoretically irrelevant.

In contrast to the above views, De Groot (2014) suggested that it is necessary to test the joint studywise hypothesis in order to test "the value of the research as a whole" (p. 189). From this perspective, studies that have a high studywise error rate have a correspondingly low research value, because their significant results are more likely to represent Type I errors. However, this reasoning assumes that the value of the research is associated with the joint studywise hypothesis, and this assumption is unwarranted unless the joint studywise hypothesis is relevant to the research question. Again, in many cases, the joint studywise hypothesis has no relevance to researchers' specific research questions, because its constituent hypotheses refer to comparisons and variables that have no theoretical or practical basis for joint consideration (Bender & Lange, 2001, p. 343; Cook & Farewell, 1996, pp. 101-102; Hewes, 2003, p. 450; Morgan, 2007, p. 34; Parker & Weir, 2020, p. 2; Perneger, 1998, p. 1236; Rothman et al., 2008, pp. 236-237; Rubin, 2020, 2022; Savitz & Olshan, 1995, p. 905; Schulz & Grimes, 2005, p. 1592). They are what Meehl (1978, p. 813) might call "a mere conjunction of unrelated assertions, a 'heap of hypotheses'." For example, in a study of alcohol and drug use disorders among homeless veterans, researchers used a Bonferroni correction when testing differences across a diverse range of variables, including age, gender, race, marital status, housing status, and mental health diagnoses (Tsai et al., 2014). In this case, it is unclear how a single joint alternative hypothesis might explain differences on all of these variables, and the researchers did not attempt this type of explanation. Consequently, it is unclear why it was necessary to adjust the alpha level on the basis of a studywise family of tests. Rothman et al. (2008, pp. 236-237) noted a similar problem in the field of epidemiology:

> A large health survey or cohort study may collect data pertaining to many possible associations, including data on diet and cancer, on exercise and heart disease, and perhaps many other distinct topics. A researcher can legitimately deny interest in any joint hypothesis regarding all of these diverse topics, instead wanting to focus on those few (or even one) pertinent to his or her specialities. In such situations, multiple-inference procedures…are irrelevant, inappropriate, and wasteful of information.

In general then, researchers should not be concerned about erroneous answers to questions that they are not asking. In other words, they should not be concerned about the familywise error rate for a joint studywise null hypothesis that they are not, in fact, testing. Instead, they should be concerned about the error rates for the individual and/or joint hypotheses about which they actually make inferences (Cook & Farewell, 1996, p. 107; Cox, 1965, p. 223; Hochberg & Tamrane, 1987, p. 6).

In some cases, the joint studywise hypothesis may subsume a collection of hypotheses that are all derived from the same theory. In this case, researchers may want to test the joint studywise hypothesis in order to make a statistical inference about the theory. However, as explained above, it is more appropriate to use conjunction testing, rather than disjunction testing, to test theories. Conjunction testing does not require an alpha adjustment. However, it may suffer from low power.

The assumption that studywise error rates should be considered on an automatic basis also forms part of an argument against the use of significance testing in exploratory research situations and in favour of the preregistration of analysis plans (e.g., de Groot, 2014; Forstmeier et al., 2017; Nosek et al., 2019, p. 816; Nosek et al., 2018; Nosek & Lakens, 2014). According to this argument, the number of hypotheses that are tested or could be tested in exploratory research situations is unknown. Consequently, the size of the family of hypotheses that comprise the joint studywise hypothesis is unknown, and an appropriate alpha adjustment cannot be computed to control the associated studywise error rate (Hochberg & Tamrane, 1987, p. 6). Again, this argument assumes that researchers are interested in disjunction testing a joint studywise hypothesis that includes all of the constituent hypotheses that they tested or could have tested in other instances of their exploratory study. However, if researchers are not interested in a disjunction test of this joint studywise hypothesis, then it becomes unnecessary for them to preregister their tests in order to control the associated studywise error rate and the vague atheoretical probability statement that this error rate underwrites (e.g., "our study yielded a significant effect, $p < .05$"). Instead, it is sufficient for researchers to make their research materials and data set publicly available (e.g., via the Open Science Framework https://osf.io/) in order for their audience to confirm that any joint hypothesis that they disjunction tested includes all of the relevant constituent hypotheses (Rubin, 2017b, pp. 272-273; Rubin, 2020, 2022). Note that, in this case, although the exploratory, post hoc disjunction testing of a series of different joint hypotheses will inflate the error rate for the (usually irrelevant) joint studywise hypothesis, it will not inflate the error rates for each of the specific, theoretically informative joint hypotheses because, by definition, each error rate is limited to the constituent hypotheses within each joint hypothesis.

Finally, the automatic consideration of studywise error rates also forms the basis for the recommendation to limit the number of tests that are performed in any given study (e.g., Armstrong, 2014; Cohen, 1990; Drachman, 2012; Goeman & Solari, 2014; Luck & Gaspelin, 2017; Schochet, 2009; Schulz & Grimes, 2005; Senn, 2007, p. 150; for a review, see Frane, 2015; for a discussion, see Wilson, 1962, p. 299). For example, in his article on multiple testing in social policy impact evaluations, Schochet (2009) advised that "limiting the number of outcomes and subgroups…is one of the best ways to address the multiple comparisons problem" (p. 548). Similarly, in their article on multiple comparison corrections in ophthalmology research, Stacey et al. (2012) suggested that "the best way to address the problem is to limit the number of comparisons" (p. 1830). Again, if researchers undertake disjunction testing of a joint hypothesis that relates to all of the variables in their study, and they do not adjust their $α_{Constituent}$ alpha level, then the more variables that they include in their study, the greater the probability that they will make a Type I error with respect to the joint studywise hypothesis. However, this issue should not deter researchers from including relevant outcome variables in their study and then adjusting their alpha level accordingly. In addition, this issue assumes that all of the outcome variables in a study relate to the same joint hypothesis and, as discussed above, this may not be the case. Finally, the number of outcome variables in a study has no impact on alpha levels that are associated with either individual testing or conjunction testing (although increasing the number of variables would decrease the power of conjunction tests). Hence, in some cases, limiting the number of tests that are conducted in a study is unnecessary.

In summary, the usefulness of studywise error rates depends on the theoretical and/or practical relevance of the joint studywise hypothesis. If this joint hypothesis is relevant to the research questions under consideration, then researchers should test it, and if they undertake a disjunction test, then they should adjust their alpha level. However, if the joint studywise hypothesis is irrelevant, then it should not be tested, and a corresponding alpha adjustment is not required (Cook & Farewell, 1996, p. 107; Cox, 1965, p. 223; Rothman et al., 2008, pp. 236-237; Savitz & Olshan, 1995, p. 905; Wilson, 1962). Furthermore, if conjunction testing is used to test the joint studywise hypothesis, then no alpha adjustment is required. Under these latter two conditions, it is inappropriate to "count the number of tests reported in a paper and multiply it by .05 to get a rough idea of the number of P values less than .05 that would be expected to arise by chance alone (if no effects being tested were real)" (Sainani, 2009, p. 1101).

*When to Adjust Alpha* 21**Multiway ANOVAwise Error Rates**

The automatic alpha adjustment assumption applies to not only large families of hypotheses, such as those that comprise a joint studywise hypothesis, but also smaller families of hypotheses, such as those tested in a multiway ANOVA or multiple linear regressions (for the same comparison, see Yekutieli et al., 2006, p. 416) Hence, some researchers believe that it is necessary to control the multiway ANOVAwise error rate (e.g., Cramer et al., 2016; Kromrey & Dickinson, 1995; Luck & Gaspelin, 2017; Rodriguez, 1997; for a more moderate positions, see An et al., 2013; Kozak & Powers, 2017).

Consider the example that Cramer et al. (2016) used to argue that alpha adjustment is necessary in exploratory multiway ANOVAs. Cramer et al. discussed a 2 (speed-stress: high/low) x 3 (age: 14-20 yrs/50-60 yrs/75-85 yrs) ANOVA that was conducted on response time data. This ANOVA tests three hypotheses: (a) a main effect of speed-stress, (b) a main effect of age, and (c) an interaction between speed-stress and age. Cramer et al. argued that "the multiway ANOVA brings with it the problem of multiple comparisons" (p. 640), because these three null hypotheses form a joint null hypothesis. As they explained,

> in an exploratory setting, all hypotheses implied by the design are considered and tested jointly, rendering this collection of hypotheses a family; in line with the idea that "the term 'family' refers to the collection of hypotheses…that is being considered for joint testing" (Lehmann & Romano, 2005). As a result, we argue that a multiple comparison problem lurks in these exploratory uses of a multiway ANOVA (p. 641).

Certainly, in an exploratory setting, it is likely that researchers would be interested in testing all three hypotheses in this multiway ANOVA. However, in an exploratory setting, it is also likely that researchers would not have any clear theoretical or practical reason for subsuming these three hypotheses under a joint ANOVAwise hypothesis and making a statistical inference based on disjunction tests of this hypothesis. Consequently, in this particular example, it is unlikely that researchers would want to adjust their alpha level for each hypothesis in order to control the multiway ANOVAwise error rate. Instead, it is more likely that they would use an individual testing approach and test each of the three hypotheses (i.e., the two main effects and the interaction effect) at their own individual, unadjusted alpha levels (i.e., $\alpha_{Individual}$).

But is it *ever* necessary to adjust alpha in order to compensate for multiple testing in multiway ANOVAs? One reason why researchers might consider adjusting their alpha levels in this context is if the ANOVA tested a group of hypotheses that were all predicted by the same theory. In this case, the researchers might want to undertake a test of that theory in the form of a joint hypothesis. However, as explained above, it is more appropriate to use an "all-tests-significant" conjunction approach for theory testing than it is to use an "at-least-one-test-significant" disjunction approach, and conjunction testing does not require an alpha adjustment. Hence, it is unlikely that researchers would ever have reasonable grounds for adjusting their alpha to compensate for multiple testing in a multiway ANOVA.[9]

To illustrate, it is useful to consider the type of inference that might be made after correcting for multiple testing in Cramer et al.'s (2016) multiway ANOVA example. Specifically, imagine that a group of researchers adjusted their alpha level from .05 to .015 in order to compensate for the three multiple tests that they conducted (i.e., the two main effects and the interaction effect). Further imagine that, using this adjusted alpha level, the researchers found a significant main effect of speed-stress but no significant effect of age and no significant speed-stress by age interaction effect. Following a disjunction decision rule, the significant speed-stress main effect would be sufficient grounds to warrant the rejection of the joint null hypothesis that "*neither* speed-stress *nor* age *nor* their interaction are related to response times." Logically, this is a correct statistical inference, because the significant speed-stress main effect refutes this joint null hypothesis (Hewes, 2003). However, it is not an inference that researchers are likely to be interested in making unless a theory predicts that "*either* speed-stress *or* age *or* their interaction are related to response times." Scientific theories do not usually specify a disjunction relation between their predictions. Instead, it is more likely that a theory would predict that "speed-stress *and* age *and* both in combination are related to response times." Consequently, it would be more appropriate to test this joint hypothesis using conjunction testing.

## Conclusions

The multiple testing literature provides plenty of advice about *how* to adjust alpha levels, but it is relatively silent about *when* to adjust alpha levels. Some previous work in this area has suggested that alpha adjustment is only necessary in exploratory research situations (e.g., Armstrong, 2014; Cramer et al., 2016; Streiner, 2015; for a review, see Frane, 2015), whereas other work has suggested that alpha adjustment is only necessary in confirmatory research situations (e.g., Bender & Lange, 2001; Schochet, 2009; Stacey et al., 2012; Tutzauer, 2003; Wason et al., 2014). In this present paper, I argued that this focus on exploratory versus confirmatory research settings is misleading, and that what really matters is the type of multiple testing that is employed: disjunction testing, conjunction testing, or individual testing.

If researchers make a decision about a joint null hypothesis after rejecting *at least one* (and not all) constituent null hypotheses, then an alpha adjustment is necessary. This disjunction testing approach is most useful when researchers aim to test a joint hypothesis without demonstrating the extent of generalisation across constituent hypotheses.

In contrast, if researchers make a decision about a joint null hypothesis after rejecting *all* of its constituent null hypotheses, then no alpha adjustment is necessary. This conjunction testing approach is most useful when all of the constituent hypotheses need to be confirmed in order to confirm the joint hypothesis.

Finally, if researchers make a decision about each null hypothesis separately, and they do not make a decision about joint null hypotheses, then no alpha adjustment is needed. Nonetheless, researchers should carefully consider the way in which they specify their alpha level during individual testing, and they should specify a lower alpha level when more stringent testing is required.

The above qualifications and limitations make it inappropriate for researchers to automatically assume that alpha adjustment is necessary in the context of multiple testing. In particular, researchers should be cautious about applying default corrections for multiple testing in relation to studywise and multiway ANOVAwise families of hypotheses.

## Endnotes

1. In the Neyman-Pearson approach, some researchers may consider *alpha size* tests rather than *alpha level* tests (Casella & Berger, 2002). However, alpha size tests are difficult to construct in the case of disjunction and conjunction testing (Casella & Berger, 2002, p. 385). Consequently, I refer to alpha level tests here.

2. The researchers could also collapse the green and red jelly beans conditions together and compare jelly beans versus the control (sugar pill) group, but they could do so on two measures of acne (e.g., inflammatory and noninflammatory). In this case, the researchers would be undertaking two tests of the same null hypothesis using two different outcome variables or *endpoints*. To keep things simple, I refer to the multiple comparisons example throughout this article. However, my arguments are equally applicable to the multiple endpoints situation.

3. The familywise error rate assumes that test results are independent. As Greenland (2020, p. 17) explained, the term *independence* is used to refer to several different concepts. In particular, he distinguished between *logical* and *statistical* independence. Logical independence refers to the mathematical independence of parameter values such that variation in one value is not logically dependent on variation in another. Logical independence may be demonstrated via the mathematics of a model. Statistical independence refers to independence among variables, estimators, standard errors, and tests, and it may be achieved via study design (e.g., randomisation). A weak form of statistical independence is *uncorrelatedness*, which assumes that there is no monotonic linear association between the variables (e.g., no positive correlation). As Greenland noted, "uncorrelatedness and hence statistical independence are rarely satisfied in nonexperimental studies." Although this may be the case, two points allow a qualified interpretation of the familywise error rate under the assumption of independence. First, when interpreting the results of a disjunction test, researchers may adopt a counterfactual interpretation that (a) the joint null hypothesis is true and (b) all of the associated test assumptions are true, including the assumption of independence. Second, researchers may complement this qualified interpretation with an acknowledgment that, if the constituent test results were positively dependent, then the actual familywise error rate would be *less* than the nominal familywise error rate, because a family of dependent tests provides less opportunity to incorrectly reject the joint null hypothesis than a family of independent tests (e.g., Weber, 2007, p. 284). Hence, although the assumption of independence may not be met in reality, researchers may nonetheless interpret the familywise error rate as indicating a worst-case scenario that assumes that the constituent test results are independent.

4. Instead of adjusting their alpha level downwards, researchers can adjust their *p* values upwards (e.g., Pan, 2013; Westfall & Young, 1993). However, there are reasons to prefer alpha adjustment over *p* value adjustment (van der Zee, 2017).

5. Some commentators have argued that conjunction testing *decreases* the Type I error rate and therefore warrants a corresponding *increase* in the $\alpha_{Constituent}$ level above the $\alpha_{Joint}$ level (e.g., Capizzi & Zhang, 1996; Massaro, 2009; Weber, 2007). This argument is based on the assumption that the Type I error rate for *k* independent tests is the product of the Type I error rate for each test (i.e., $\alpha^k$). Hence, for example, the probability of obtaining two independent false positive results at the .05 alpha level is only .0025. However, during conjunction testing, *all* of the tests are required to be significant in order to reject the joint null hypothesis. Consequently, when undertaking conjunction testing, the alpha level for each of the constituent null hypotheses ($\alpha_{Constituent}$) cannot be higher than the alpha level for the joint null hypothesis ($\alpha_{Joint}$; Berger, 1982; Julious & McIntyre, 2012; Kordzakhia et al., 2010).

6. Tukey (1953), who was a pioneer in the area of multiple testing, described this individual testing error rate as the *per determination* error rate (i.e., $\alpha_{Individual}$). This error rate should not be confused with the *per comparison* error rate (i.e., $\alpha_{Constituent}$). Both error rates use unadjusted alpha levels. However, the per determination error rate is used in the context of the individual testing of an individual null hypothesis, whereas the per comparison error rate is used in the context of the disjunction testing of a joint null hypothesis. Tukey (p. 90) was firmly against the use of the per comparison error rate. However, he believed that the per determination error rate was "entirely appropriate" (p. 82) for some research questions (i.e., individual testing; see also Hochberg & Tamhane, 1987, p. 6). For example, he argued that a per determination rate was suitable when diagnosing potentially diabetic patients based on their blood sugar levels. As Tukey (1953, p. 82) explained:

   the doctor's action on John Jones would not depend on the other 19 determinations made at the same time by the same technician or on the other 47 determinations on samples from patients in Smithville. Each determination is an individual matter, and it is appropriate to set error rates accordingly.

7. A selection bias remains problematic during individual testing, because it involves the *suppression of hypotheses after the results are known* or SHARKing (Rubin, 2017d). SHARKing is problematic when suppressed falsifications are theoretically (as opposed to statistically) relevant to the research conclusions. For example, in the jelly bean study, it is theoretically informative to know not only that green jelly beans cause acne but also that non-green jelly beans do not appear to cause acne.

8. Studywise and multiway ANOVAwise error rates are not the only types of error rates that have caused confusion in the area of multiple testing. Other examples include *datasetwise* error rates (in which the family includes all hypotheses that are tested using a specific dataset; Bennett et al., 2009, p. 417; Thompson et al., 2020), *careerwise* error rates (in which the family includes all hypotheses that are performed by a specific researcher during their career; O'Keefe, 2003; Stewart-Oaten, 1995), and *fieldwise* error rates (in which the family includes all hypotheses that are performed in a specific field). A key argument in the current article is that researchers do not usually make decisions about data sets, researchers, and fields. Instead, they make decisions about hypotheses.

9. Multiple testing corrections may be necessary in multiway ANOVAs when a factor contains more than two levels and multiple comparisons are conducted between those levels in order to test a joint intersection null hypothesis (Benjamini & Bogomolov, 2011; Yekutieli et al., 2006). However, in this case, familywise error rates are limited to the comparisons that are made within factors. Familywise error is not computed across all factors in the ANOVA.

## Funding
The author declares no funding sources.

## Conflict of Interest
The author declares no conflict of interest.